\documentclass[preprint,5p,compress]{elsarticle}
\usepackage{amssymb,bm,mathrsfs,bbm,amscd}
\usepackage{mdwlist}

\usepackage{graphics}
\usepackage{xspace}
\usepackage{amsmath}
\usepackage{amsfonts}
\usepackage{srcltx}
\usepackage{hyperref}
\usepackage{psfrag}
\usepackage{epsfig}

\renewcommand{\epsilon}{\varepsilon}

\journal{Physics Letters B}

\begin{document}

\begin{frontmatter}

  \title{Search for narrow resonances in $e^+e^-$ annihilation between 
1.85 and 3.1 GeV with the KEDR Detector
  }


\author[BINP]{V.V.~Anashin}
\author[BINP,NSU]{V.M.~Aulchenko}
\author[BINP,NSU]{E.M.~Baldin}
\author[BINP]{A.K.~Barladyan}
\author[BINP]{A.Yu.~Barnyakov}
\author[BINP]{M.Yu.~Barnyakov}
\author[BINP,NSU]{S.E.~Baru}
\author[BINP]{I.Yu.~Basok}
\author[BINP,NSU]{O.L.~Beloborodova}
\author[BINP]{A.E.~Blinov\corref{1}}
\cortext[1]{Corresponding author:A.E.Blinov@inp.nsk.su}
\author[BINP,NSTU]{V.E.~Blinov}
\author[BINP]{A.V.~Bobrov}
\author[BINP]{V.S.~Bobrovnikov}
\author[BINP,NSU]{A.V.~Bogomyagkov}
\author[BINP,NSU]{A.E.~Bondar}
\author[BINP]{A.R.~Buzykaev}
\author[BINP,NSU]{S.I.~Eidelman}
\author[BINP]{D.,N.~Grigoriev}
\author[BINP]{Yu.M.~Glukhovchenko}
\author[BINP]{V.V.~Gulevich}
\author[BINP]{D.V.~Gusev}
\author[BINP]{S.E.~Karnaev}
\author[BINP]{G.V.~Karpov}
\author[BINP]{S.V.~Karpov}
\author[BINP,NSU]{T.A.~Kharlamova}
\author[BINP]{V.A.~Kiselev}
\author[BINP]{V.V.~Kolmogorov}
\author[BINP,NSU]{S.A.~Kononov}
\author[BINP]{K.Yu.~Kotov}
\author[BINP,NSU]{E.A.~Kravchenko}
\author[BINP]{V.N.~Kudryavtsev}
\author[BINP,NSU]{V.F.~Kulikov}
\author[BINP,NSTU]{G.Ya.~Kurkin}
\author[BINP,NSU]{E.A.~Kuper}
\author[BINP,NSTU]{E.B.~Levichev}
\author[BINP]{D.A.~Maksimov}
\author[BINP]{V.M.~Malyshev}
\author[BINP]{A.L.~Maslennikov}
\author[BINP,NSU]{A.S.~Medvedko}
\author[BINP,NSU]{O.I.~Meshkov}
\author[BINP]{S.I.~Mishnev}
\author[BINP,NSU]{I.I.~Morozov}
\author[BINP,NSU]{N.Yu.~Muchnoi}
\author[BINP]{V.V.~Neufeld}
\author[BINP]{S.A.~Nikitin}
\author[BINP,NSU]{I.B.~Nikolaev}
\author[BINP]{I.N.~Okunev}
\author[BINP,NSTU]{A.P.~Onuchin}
\author[BINP]{S.B.~Oreshkin}
\author[BINP,NSU]{I.O.~Orlov}
\author[BINP]{A.A.~Osipov}
\author[BINP]{S.V.~Peleganchuk}
\author[BINP,NSTU]{S.G.~Pivovarov}
\author[BINP]{P.A.~Piminov}
\author[BINP]{V.V.~Petrov}
\author[BINP]{A.O.~Poluektov}
\author[BINP]{V.G.~Prisekin}
\author[BINP]{A.A.~Ruban}
\author[BINP]{V.K.~Sandyrev}
\author[BINP]{G.A.~Savinov}
\author[BINP]{A.G.~Shamov}
\author[BINP]{D.N.~Shatilov}
\author[BINP,NSU]{B.A.~Shwartz}
\author[BINP]{E.A.~Simonov}
\author[BINP]{S.V.~Sinyatkin}
\author[BINP]{A.N.~Skrinsky}
\author[BINP,NSU]{V.V.~Smaluk}
\author[BINP]{A.V.~Sokolov}
\author[BINP]{A.M.~Sukharev}
\author[BINP,NSU]{E.V.~Starostina}
\author[BINP,NSU]{A.A.~Talyshev}
\author[BINP]{V.A.~Tayursky}
\author[BINP,NSU]{V.I.~Telnov}
\author[BINP,NSU]{Yu.A.~Tikhonov}
\author[BINP,NSU]{K.Yu.~Todyshev}
\author[BINP]{G.M.~Tumaikin}
\author[BINP]{Yu.V.~Usov}
\author[BINP]{A.I.~Vorobiov}
\author[BINP]{A.N.~Yushkov}
\author[BINP]{V.N.~Zhilich}
\author[BINP,NSU]{V.V.~Zhulanov}
\author[BINP,NSU]{A.N.~Zhuravlev}

\address[BINP]{Budker Institute of Nuclear Physics SB RAS, 
Novosibirsk, 630090, Russia}
\address[NSU]{Novosibirsk State University, Novosibirsk, 630090, Russia}
\address[NSTU]{Novosibirsk State Technical University, Novosibirsk, 
630092, Russia}

\begin{abstract}

We report results of a search for narrow resonances in $e^+e^-$ 
annihilation at center-of-mass energies between 1.85 and 3.1 GeV 
performed with the KEDR detector at the VEPP-4M $e^+e^-$ collider. 
The upper limit on the leptonic width of a narrow resonance 
$\Gamma^R_{ee}\cdot Br(R\to hadr) < 120$ eV has been obtained 
(at 90\% C.L.).

\end{abstract}


\end{frontmatter}


\section{Introduction}
\label{sec:intro}
After the  \(J/\psi\) discovery, a search for other narrow resonances 
was performed in several experiments. 
The energy range between \(J/\psi\) and $\Upsilon$ mesons was explored with MARK-1 at SPEAR~\cite{siegrist-1982}, LENA at DORIS~\cite{nickzuporuk-1982}, and MD-1 at VEPP-4~\cite{blinov-1991}. 
The upper limit on the leptonic width of narrow resonances obtained 
in these analyses varied between 15 and 970 eV depending on energy. 
Searches in the energy range below \(J/\psi\) mass and down to 1.42 GeV 
were performed only in experiments at ADONE~\cite{esposito-1975,
baldini-1975,baldini-1976,ambrosio-1976,esposito-1976,
baldini-1978,ambrosio-1978} with upper limits of about 500 eV. 
Recently the latter energy region was revisited by the KEDR
collaboration~\cite{Blinov2010} in view 
of a discovery of unexpected exotic states above the charm 
threshold, including the narrow $X(3872)$ state, which proved that 
surprises are still possible even at low energies~\cite{nora}.
This paper reports results of a search for narrow resonances in 
$e^+e^-$ annihilation in the center-of-mass (c.m.) energy range 
1.85-3.1 GeV. The experiment was performed at the VEPP-4M $e^+e^-$
collider in Novosibirsk in 2009 and 2010.

The outline of this paper is as follows. 
In Section 2 we describe our apparatus and trigger conditions. 
Section 3 describes the experiment. 
Section 4 deals with hadronic event selection.
Section 5 covers a fit procedure and results while Section 6 is reserved 
for conclusions.

\section{VEPP-4M collider and KEDR detector}
\label{sec:VEPPKEDR}

The electron-positron accelerator complex VEPP-4M~\cite{Anashin:1998sj} 
is designed for high-energy physics experiments in the c.m. energy 
range ($W$) from 2 to 12 GeV.
The collider consists of two half-rings, an experimental section where
the KEDR detector is installed, and a straight section, which includes an
RF cavity and injection system.
The luminosity at the $J/\psi$ in an operation mode with 2 by 2 bunches 
reaches~$1.5\times10^{30}\,\text{cm}^{-2}\text{s}^{-1}$.


One of the main features of the VEPP-4M is its capability to precisely 
measure beam energy using two techniques~\cite{Blinov:2009}: 
resonant depolarization and infrared light Compton backscattering.
The accuracy of VEPP-4M energy determination with  resonant 
depolarization reaches  $\simeq$10~keV 
in the \(J/\psi\) region~\cite{Aulchenko:2003qq}.
However, such measurement is time-consuming and requires dedicated 
calibration runs without data taking. 
A new technique developed at the BESSY-I and BESSY-II 
synchrotron radiation sources~\cite{Klein:1997wq,Klein:2002ky} was 
adopted for VEPP-4M in 2005. It employs the infrared light Compton 
backscattering and has worse precision compared to the resonant 
depolarization~(50$\div$70~keV in the \(J/\psi\) region), 
but unlike the latter can be used during data taking~\cite{Blinov:2009}.

\begin{figure}[t]
\begin{center}
\vspace{-9mm}
\includegraphics[width=\columnwidth]{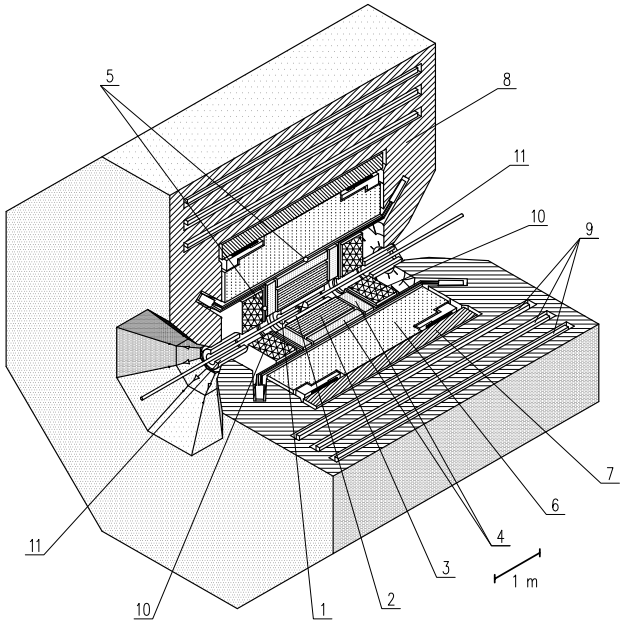}
\vspace{-32mm}
\caption{\label{fig:KEDR} KEDR detector. 1-vacuum chamber, 2-vertex detector, 
3-drift chamber, 4-threshold aerogel counters, 
5-ToF-counters, 6-liquid krypton calorimeter, 7-superconducting coil (0.6 T), 
8-magnet yoke, 9-muon tubes, 10-CsI-calorimeter,
11-compensation solenoid.}
\end{center}
\end{figure}

The KEDR detector (Fig.~\ref{fig:KEDR}) is described in detail 
elsewhere~\cite{Anashin:2002uj}. 
It  includes a tracking system consisting of a vertex detector and 
a drift chamber, a particle identification system of aerogel 
Cherenkov counters and scintillation time-of-flight counters, 
and an electromagnetic calorimeter based on liquid krypton 
in the barrel and CsI crystals in the endcap. 
The superconducting solenoid provides a longitudinal magnetic field of 0.6~T.
A muon system is installed inside the magnet yoke. 
The detector also includes a high-resolution tagging system for 
studies of two-photon processes.
The online luminosity measurement is performed with two sampling 
calorimeters which detect photons from the process of single brehmsstrahlung.

A trigger of the KEDR detector consists of two levels: primary and secondary. 
Both operate at the hardware level and are designed to provide 
high efficiency for events with two charged tracks. 
A primary trigger uses signals from the time-of-flight counters and 
calorimeters as inputs, and the typical rate is $5\div 10$ kHz.
It can be fired by one of the following conditions: two separated 
time-of-flight counters or one time-of-flight counter and one 
cluster in the endcap calorimeter or two clusters in the opposite 
endcap calorimeters or one cluster in the barrel calorimeter.
A secondary trigger uses signals from the vertex detector, drift chamber, 
and muon system in addition to the systems listed above, and the rate is 
$50\div150$ Hz.

\section{Experiment description}
\label{sec:Experiment}

\begin{figure}
\begin{center}
\vspace{-10mm}
\includegraphics[width=\columnwidth]{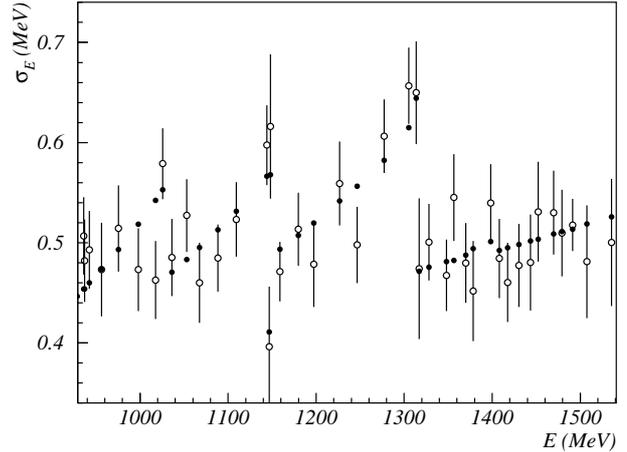}
\end{center}
\vspace{-8mm}
  \caption{\label{fig:sepict}  The beam energy spread 
$\sigma_E(E_{beam})$ dependence on the beam energy. 
Open circles - values of $\sigma_E$, obtained from $\sigma_z$ 
of the interaction region. Black points - fit of $\sigma_E(E_{beam})$ 
based on the currents in major magnetic elements}
\end{figure}

The experiment was performed in 2009 and 2010.
The energy scan started just above the \(J/\psi\) and 
finished at $W$= 1.85 GeV. 
The beam energy $E_{beam}$ was measured by the Compton backscattering 
technique described in Section~\ref{sec:VEPPKEDR}.
In order to maximize luminosity the magnetic structure of VEPP-4M 
was retuned with decrease of the beam energy. 
It caused non-trivial dependence of the beam energy spread $\sigma_E$ 
on $E_{beam}$.
The $\sigma_E$ was measured using its relation with 
a longitudinal size of the interaction region $\sigma_z$.  
The values of $\sigma_E(E_{beam})$ and their fit, based on the currents in 
major magnetic elements, are shown in Fig.~\ref{fig:sepict}.

A search for narrow resonances was conducted by automatic decrease 
of the c.m. energy by about $2\sigma_W$ (1.4 to 1.9 MeV) steps 
after collection of required integrated luminosity at each point.
In order to get energy-independent sensitivity in terms of 
$\Gamma^R_{ee}$, the integrated luminosity per energy point 
varied from 0.3~nb\(^{-1}\) in the upper part of the energy range to 
0.12~nb\(^{-1}\) in the lower one.
The data taken at each energy were analyzed on line. 
In order to improve sensitivity, the integrated luminosity was doubled at 
the energy points with significant excess of candidate events.

The luminosity was monitored using the process of single bremsstrahlung, 
while the analysis uses the offline measurement based on 
elastic $e^+e^-$ scattering in the endcap calorimeter.
The total integrated luminosity  $\int\! L dt \simeq$ 300~nb$^{-1}$ 
was collected.

\section{Event Selection}
\label{sec:EvSel}



The event selection has two stages.
At the first stage we define the following track-level criteria for 
charged tracks:
\begin{enumerate*}
\item The distance of the track closest approach to the beam 
in the transverse plane and along the beam axis should be less 
than 0.5 and 10 cm, respectively;
\item The energy deposited by the track in the barrel liquid 
krypton calorimeter is above 20 MeV.
\end{enumerate*}

At the second stage the following event-selection criteria, 
which take into account both tracking and calorimetric information, 
were applied:
\begin{enumerate*}
\item The total energy deposited in the calorimeter is above 400 MeV;
\item At least two charged tracks in the event satisfy first track-level 
criteria;
\item At least one ``good'' charged track satisfies both track-level criteria;
\item There is a charged track acoplanar to the ``good'' one: 
$|\Delta\phi-\pi|>$0.15~rad;
\item Aplanarity of event (a sum of momenta transverse to 
the ``event plane'') is above $0.1~E_{beam}$;
\item There are less than 4 hits in the muon chambers;
\item $|\Sigma p_z/\Sigma E|<0.5$, where $\Sigma p_z$ and $\Sigma E$ are the total longitudional momentum and energy of all particles, respectively.
\end{enumerate*}

Condition 4 rejects cosmic rays, Bhabha, and dimuon events. 
Condition 5 rejects radiative Bhabha events and dimuons. 
Condition 6 rejects cosmic ray showers.
Condition 7 rejects two-photon processes and events with hard 
initial state radiation.
\section{Fit procedure and results}
\label{sec:Fit}


The observed number of hadronic events in data taking runs 
($N_{i}^{\mathrm{obs}}$) was fitted
using the maximum likelihood method \cite{Baker}:
\begin{equation}
\begin{split}
-2 \,\ln{ \mathcal{L}} & = 2 \sum_{i} \Bigg[ N_{i}^{\mathrm{obs}} 
  \ln{ \Bigg( \frac{N_{i}^{\mathrm{obs}}}{ N_{i}^{\mathrm{exp}} } \Bigg)} +  N_{i}^{\mathrm{exp}} -  N_{i}^{\mathrm{obs}} \Bigg] \,
\label{ML:func}
\end{split}
\end{equation}
The expected number of hadronic events in data taking runs 
($N_{i}^{\mathrm{exp}}$) is obtained as 
\begin{equation}
 N_{i}^{exp}  =  \sigma(W_{i}) \cdot L_{i}, \,
\label{Nexp}
\end{equation}
where $\sigma(W_{i})$ and $L_{i}$ are the  hadronic cross section 
at the c.m. energy of the run and the run integrated luminosity, 
respectively. 

The hadronic cross section $\sigma(W)$ is parameterized with 
a function that assumes the existence of a resonance with 
mass $M_R$, and the leptonic width $\Gamma^R_{ee}$ on top of the flat 
non-resonant background:
\begin{equation}
\begin{split}
\label{eq:convolution}
\sigma(W)=\sigma_0+\epsilon_{h}(M_R) \int\! dW'\,&dx \cdot \\
\cdot\sigma_{e^+ e^- \to R \to hadr}(W')\cdot& \mathscr{F}(x,W') G\Big(\frac{W-W'}{\sigma_{W}(M_R)}\Big),
\end{split}
\end{equation}
where
\begin{equation}
\sigma_{e^+ e^- \to R \to hadr}(W)=\frac{6\pi^2}{M^2_R}\Gamma^R_{ee}\cdot Br(R\to hadr)\cdot\delta(W-M_R),  \nonumber
\end{equation}
$\sigma_0$ is the non-resonant background cross section 
at $W = M_R$, $\epsilon_h$ is the event selection efficiency 
for the resonance hadronic decays, $\mathscr{F}(x,W)$ is the 
radiative correction function~\cite{KuraevFadin}, and $G(x)$ is 
the Gaussian function.
The value of  $\sigma_{W}(M_R)$ is obtained from the fit 
plotted in Fig.~\ref{fig:sepict} as $\sigma_{W} = \sqrt{2}\sigma_{E}$.

In order to have enough background events for a $\sigma_0$ measurement, 
the fits use the energy range $M_R\pm 13$ MeV. 
The fits were performed with $M_R$ varied by 0.1~MeV steps taking 
only $\sigma_0$ and $\epsilon_h\Gamma^R_{ee}\cdot Br(R\to hadr)$ 
as free parameters.
The fits did not reveal statistically significant narrow resonances 
other than $J/\Psi$.

\begin{figure*}[t]
\begin{center}
\includegraphics[]{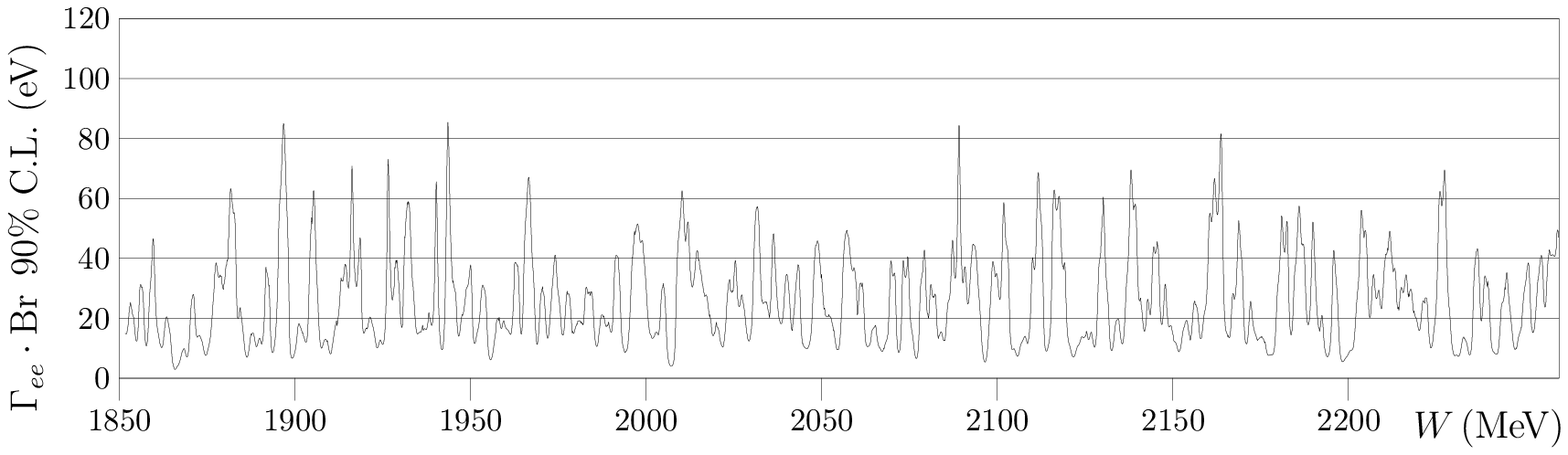}
\includegraphics[]{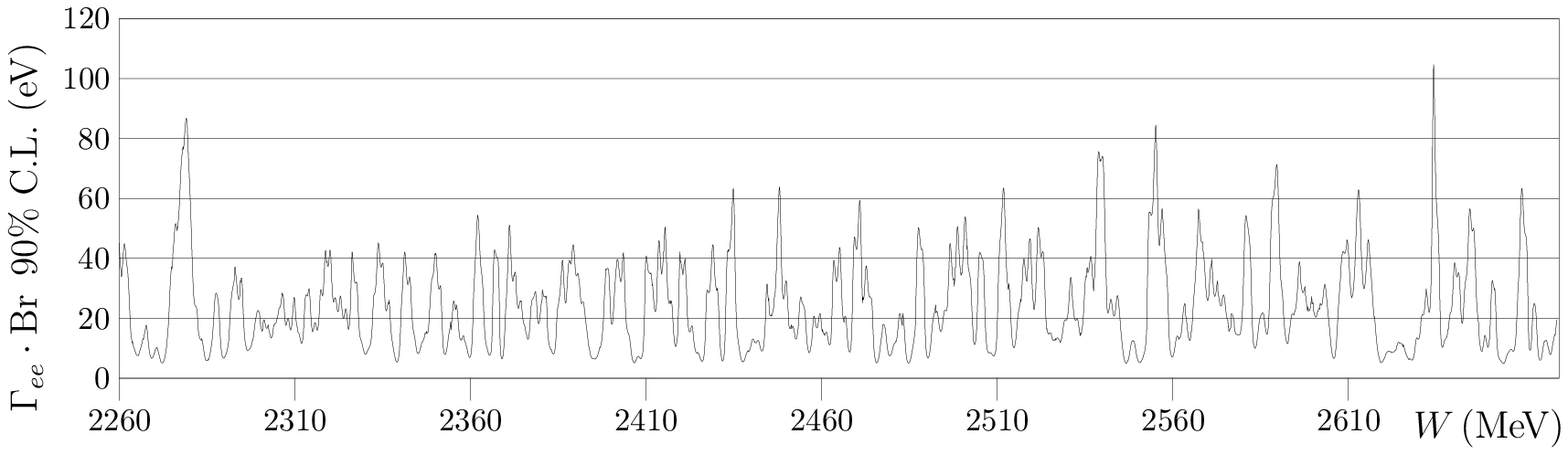}
\includegraphics[]{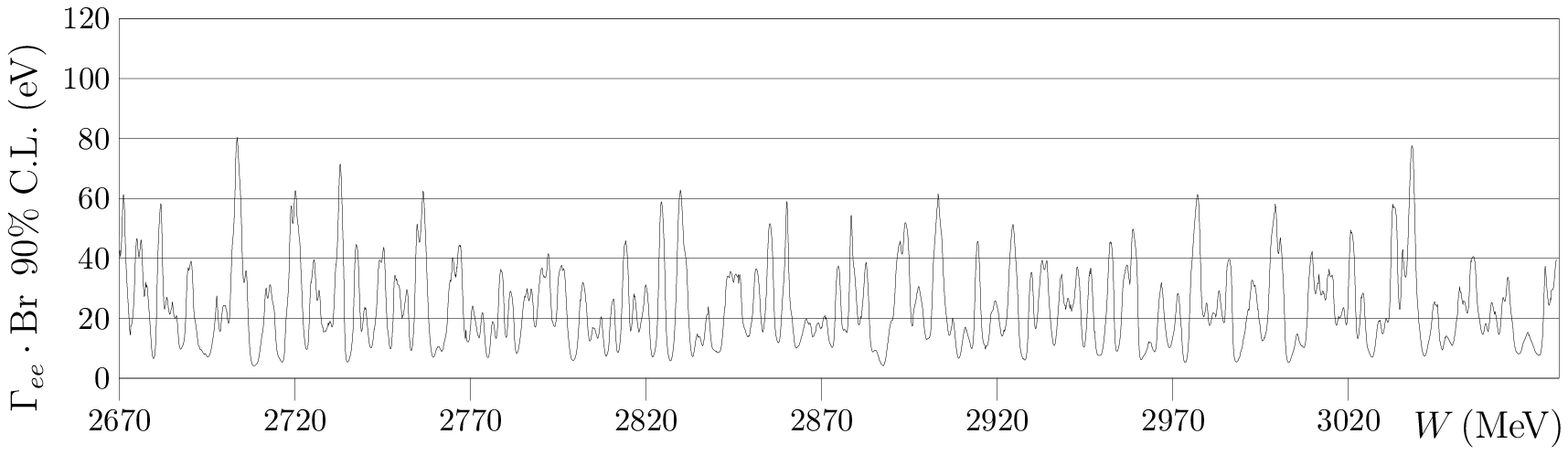}
\end{center}
\vspace{-5mm}
  \caption{\label{fig:geelim10}  W-dependence of 
90\% C.L. upper limit on  $\Gamma^R_{ee}\cdot Br(R\to hadr)$.}
\end{figure*}

 
In order to set an upper limit on $\Gamma^R_{ee}$, 
the event selection efficiency $\epsilon_h$ should be estimated.
It has been obtained as follows:
\begin{enumerate*}
\item Since $\Gamma^{J/\Psi}_{ee}\cdot Br(J/\Psi\to hadr)$ is known from 
the other measurements~\cite{PDG2010}, $\epsilon_h(J/\Psi)$ of 
about $62\%$ was obtained at the very beginning of the scan in 
the $J/\Psi$ energy range.
\item The hadronic decays of a hypotetical resonance might differ 
from those of the $J/\Psi$ one.  
The relative uncertainty of $\epsilon_h$ due to this factor was 
estimated as 10\% from comparison of two Monte Carlo simulations: 
$J/\Psi \to hadrons$ and $e^+ e^- \to hadrons$ in continuum. 
\item Variation of $\epsilon_h$ with energy was estimated from 
the drop of the visible cross section of $e^+ e^- \to hadrons$ 
after it was corrected for the $W^{-2}$ dependence.  
Taking into account R measurements in this energy region~\cite{BES2002}, 
the relative decrease of $\epsilon_h$ in the experimental 
energy range was estimated as $(22 \pm 7)\%$.
\item Taking all the factors conservatively, $\epsilon_h(W)$ 
has been obtained by the linear interpolation between 56\% at W~=~3.1 GeV 
and 40\% at W~=~1.85 GeV.
\end{enumerate*}
The energy-dependent 90\% C.L. upper limits on 
$\Gamma^{J/\Psi}_{ee}\cdot Br(J/\Psi\to hadr)$ with the highest value 
of about 105 eV, obtained from such a fit, are 
shown in Fig.~\ref{fig:geelim10}.
\footnote{Detailed information on the results of our analysis can be found in two tables in the electronic supplement to this paper}
Variation of $\sigma_{W}$ within its 10\% systematic uncertainty 
could increase the limit to 120 eV.


\section{Conclusions}
\label{sec:Conclusions}


A detailed study of the energy range 1.85-3.1 GeV at the 
VEPP-4M collider with the KEDR detector revealed no new narrow 
resonances in the reaction $e^+e^-\to hadrons$. An upper limit 
obtained for the leptonic width of possible resonances
\begin{equation}
\Gamma^R_{ee}\cdot Br(R\to hadr) < 120~{\rm eV}~(90\%~{\rm C.L.}) \nonumber
\end{equation}
is four to five times more stringent than that obtained in this 
energy range in earlier experiments at ADONE~\cite{esposito-1975,baldini-1975,baldini-1976,ambrosio-1976,esposito-1976,baldini-1978,ambrosio-1978}.


\section{Acknowledgments}
This work was partially supported by the Russian Foundation for 
Basic Research, Grants 07-02-00816-a, 10-02-00695-a, 11-02-00112-a, 
11-02-00558-a and 
RF Presidential Grant for Sc. Sch. NSh-5655.2008.2.


\vspace{-2mm}
\centerline{\rule{80mm}{0.1pt}}
\vspace{2mm}



\end{document}